\documentstyle[twocolumn,aps,psfig,floats]{revtex} 
\newcommand{\ad}{a^\dagger} 
\newcommand{\bra}[1]{\left<#1\right|} 
\newcommand{\ket}[1]{\left|#1\right>} 
\newcommand{\braket}[2]{\left<#1\right.\left|#2\right>} 
\newcommand{\eqref}[1]{(\ref{#1})}

\begin{document}
\draft
\title{Maximum efficiency of a linear-optical Bell-state analyzer}
\author{John Calsamiglia\thanks{e-mail: John.Calsamiglia@helsinki.fi} and Norbert L\"utkenhaus}
\address{Helsinki Institute of Physics, PL 9, FIN-00014 Helsinki
  University, Finland}
\date{\today}
\maketitle
\begin{abstract}
In a photonic realization of qubits the implementation of quantum logic is
rather difficult due the extremely weak interaction on the few photon
level. On the other hand, in these systems interference is available to
process the quantum states. We formalize the use of interference by the
definition of a simple class of operations which include linear optical
elements, auxiliary states and conditional operations.

We investigate an important subclass of these tools, namely linear optical
elements and auxiliary modes in the vacuum state. For this tools, we are
able to extend a previous quantitative result, a no-go theorem for
perfect Bell state analyzer on two qubits in polarization entanglement, by
a quantitative statement. We show, that within this subclass
it is not possible to discriminate unambiguously  four
equiprobable Bell states with a probability higher than 50 \%.
\end{abstract}

\narrowtext

\section{Introduction}

In the theory of quantum information processing we are working with the
abstract notion of qubits. These are two-level quantum systems on which
we can perform either individual operations or joint operations, for
example two-qubit gates like the CNOT operation. With these gates it is
possible to perform an arbitrary operation on a collection of qubits. 

In the field quantum communication we are interested in a photonic
implementation of qubits since photons are readily transported through
free space or optical fibers. However, the implementation of quantum
gates is rather difficult unless it becomes viable to map the photonic
qubits onto atomic states which can be more easily processed 
\cite{vEnk97}. From an
experimental point of view it would be ideal to process the photonic
qubits directly without the need to store them. At the level of
single photons it appears to be beyond present experimental capabilities
to use non-linear effects to perform two-qubit gate operations.

The implementation of qubits by photons opens another way to process the
represented quantum information. Since the qubits are now
indistinguishable particles, we can use interference by mixing the
photonic modes using linear optical elements 
\footnote{In this paper we will refer to  linear optical elements meaning
 \emph{passive} linear optical elements which mix linearly  the mode creation 
operators (see  Eq.~\eqref{eq:linComb}). In contrast \emph{active}
 linear optical elements  mix linearly creation 
and annihilation operators.}. In general, we are thereby leaving
the simple picture of two-level systems, but if we are interested purely
in the extraction of information via measurements on some qubits, rather
than  operations on qubits, then this is not a drawback at all. In this
paper we define a simple class of tools which is designed to capture
elements of a simple implementation of general measurement on photonic
qubits. It comprises mixing of optical modes via linear optical elements,
the use of auxiliary modes prepared in an arbitrary initial state,
the use of conditional dynamics which allows to control the evolution of
a subsystem conditioned on the measurement result on some other
subsystem and, finally, photon-counting measurements. 

The investigation of the measurements implementable with these tools is
important because of the direct relation to simple physical
implementations. Additionally, it is even very important to investigate
the power of subclasses of these tools. For example, measurements which
can be performed with auxiliary modes prepared in the vacuum state and
which do not need to make use of conditional dynamics are the easiest
measurements one can think of. It is therefore important to see what
basic class of measurements can thus be realized, and how the additional
tools, like non-vacuum auxiliary modes or conditional dynamics, extend our
capability to perform non-trivial measurements. 

An essential measurement in quantum communication is the Bell measurement
on a product of two separate qubit systems. This measurement is an 
essential tool for teleportation \cite{bennett93a}, 
dense coding \cite{bennett92},
quantum repeaters \cite{briegel}
and fault tolerant quantum computing \cite{gottesman}. We have
shown earlier
\cite{lut99} that it is not possible to perform a perfect Bell measurement
with these  tools on two qubits represented by the polarization state of a
photon. However, this leaves room for an apparatus which unambiguously
discriminates between the four Bell states with probability less than
one.\footnote{Actually, as we finish this manuscript, a work by Knill
and coworkers \cite{knill}
 presents a way to approach unit probability of success
asymptotically with a growing number of entangled auxiliary photons, 
making full use of all presented tools.} In the present paper, we
investigate the restricted problem where we make use only of vacuum state
auxiliary modes and exclude conditional dynamics. The aim is to provide
an upper bound on the success probability of a Bell
measurement, using the restricted set of tools.  Our result shows that 
it is possible for equiproble Bell states to obtain an unambiguous result
in half of the cases only, showing that the early proposed implementations
\cite{weinfurter94} are indeed optimal within this class.

\section{Description of viable measurements} 

Before  we continue we shall describe our tools more precisely. We
restrict our measurement  apparatus to  linear elements. This
means that the creation operators  of the output modes 
($\{c_{i}^\dagger\}$) area linear combination of the input modes 
($\{a_{i}^\dagger\}$) ,

\begin{equation}
    c_{i}^\dagger=\sum_{j=1}^{n} U^\dagger_{i j} a_{j}^\dagger
    \label{eq:linComb}
\end{equation}
where $U$ is a unitary  matrix.  Reck \emph{et al.} \cite{reck94a}  have shown that 
in fact any unitary mapping can be realized 
using only beamsplitters and  phase
shifters.
 
The  number of modes is not necessarily limited by the 
number of modes occupied by the input states: we can couple
to auxiliary  modes using  beam-splitters  so that the initial state of 
input and apparatus is 
described by  the   direct product of  the  Hilbert  space   of the
input states and  the initial state of the auxiliary modes. The 
auxiliary modes can be prepared in an arbitrary state with any number of 
excitations.  

All  modes  are mapped   into  output  modes, where we place
detectors. Detectors with photon number resolution
are difficult to realize. Nevertheless, since
 here we want to concentrate on the real power of linear elements we will work with ideal detectors, so that each measurement outcome corresponds to projection onto photon number states.

 Here it is important to emphasize that linear optical 
elements can provide any arbitrary unitary mapping 
only over creation operators, not over a general input state.
 Only in the 
case of one-photon states a unitary map on the creation operators translates into the same map on the the one-photon states.  In fact, making 
use of this result and of the Neumark extension \cite{neumark,peres93}, 
we can realize any generalized measurement over one-photon states
 using the tools described so far.

The  last  tool introduced  here is   the  ability to perform
conditional measurements. With  that  we  mean  that we monitor   one
selected mode while keeping the other modes in a waiting loop. Then we
can perform some linear operation on  the remaining modes depending on
the outcome of the measurement  with all the tools described  above. 

\begin{figure}[htb] 
\centerline{\psfig{figure=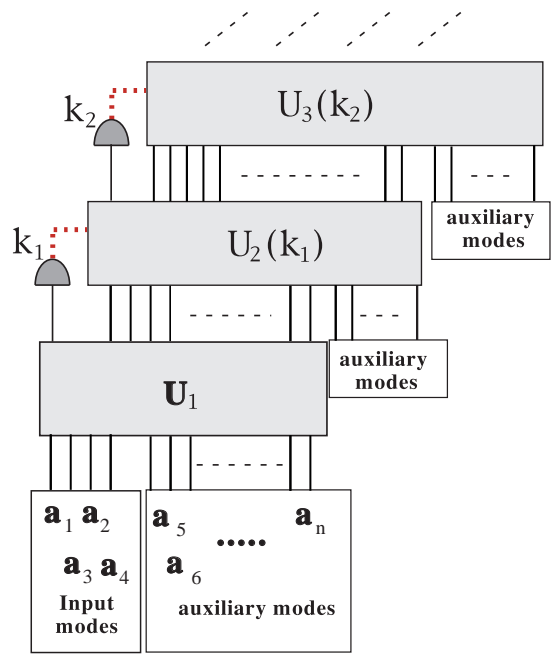,width=3in}}  
\vspace{0.2cm}
\caption{ In the general scheme the modes  of the input state are mixed with
the auxiliary modes, which are not necessarily  in  the vacuum  state.
Then  one selected mode is measured and, depending on the  measurement outcome,
the other output modes are mixed linearly with new auxiliary modes.
This procedure can be repeated indefinitely.}  
\label{fig:generalscheme} 
\end{figure}

The  general strategy  is  shown schematically  in
figure \ref{fig:generalscheme} and makes use of 
electronically switched passive linear optical elements,
auxiliary modes with or without excitations and ideal photo-detectors.

\section{Bell-state analyzer efficiency} 

A general measurement is described by a positive
operator valued measure (POVM) \cite{peres93} given by a collection of
positive operators $F_k$ with $\sum_k F_k = \openone$. Each operator 
$F_k$ corresponds to one classically distinguishable measurement 
outcome (e.g. a given combination of ``clicks'' in the output detectors).
The probability $p_k$ for the outcome
$k$ to occur while the input is being described by density matrix
$\rho$,  is given by $p_k = \mbox{Tr}(\rho F_k)$.  

A perfect Bell-state analyzer is an apparatus which performs a 
von Neumann projection  measurement on 
a maximally entangled basis. 
That is, it is an apparatus for which every measurement outcome is 
described by a POVM element proportional to a projector on one of 
the four orthogonal Bell-states.
 It has been proven already \cite{lut99} 
that with the tools described in the previous section it is not possible to 
construct  such an apparatus.

Here we will relax a bit the constraints of the perfect 
Bell-state analyzer by allowing it
to fail with some probability. 
Our non-perfect Bell-state analyzer is 
then defined as a measurement apparatus for which \emph{some} POVM 
elements $\{F_l\}$ are proportional to a projector onto one of 
the four orthogonal Bell-states. This means that the events
corresponding to each of these POVM elements can only be triggered by only 
one of the input Bell-states  ($p_{l}=\mbox{Tr}(\rho_{\Psi_i} F_l) \neq 0$
 for only one of the four Bell state
inputs $ \rho_{\Psi_i}$, $i=1,\dots,4$).
 This allows us to rephrase the problem as one of unambiguous
 state discrimination between the four Bell-states.

\begin{figure}[htb] 
\centerline{\psfig{figure=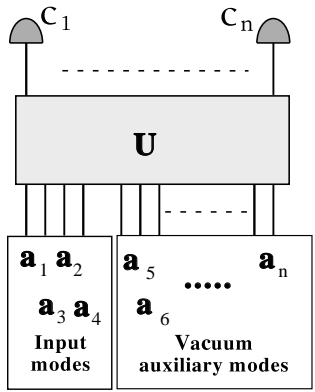,width=2in}}  
\vspace{0.2cm}
\caption{In the restricted scheme, the modes of the two-photon input states are
 linearly mixed with the auxiliary vacuum modes. Ideal detectors 
are then  placed in the output modes.}  
\label{fig:restscheme} 
\end{figure}

In this paper we determine the maximum probability of 
success for the Bell-state analyzer using a 
restricted, more basic, set of tools (see fig.
\ref{fig:restscheme}). Our 
apparatus consists of a fix array of optical elements (no conditional
measurements allowed) and  the  auxiliary modes are initially 
in the vacuum state (no extra photons allowed). 
In the remaining of the paper we will proof that for equiprobable input
 Bell-states the maximal 
probability of success of this Bell state analyzer is 
$P^s=\frac{1}{2}$.\footnote{This result can be easily generalized to 
the case where all Bell-states have a non-zero a priori probability 
but they are not equiprobable (i.e. $p_{1}\geq\ldots\geq p_{4} \neq 0 $)
giving an optimum success probability of  $P^s=p_{1}+p_{2}$.}

We can summarize the proof as follows: We introduce the 
formalism which we will use along the proof. We prove that all two-photon 
detection events are a failure (they do not contribute to the success 
probability). Then we put an upper bound on the probability that 
an output mode  $c_{i}$ is involved in a successful discrimination 
event. In order to fix this upper bound we first show that one mode can 
participate in the discrimination of at most two Bell states. 
In the end we obtain the upper bound on the total probability of 
success by adding the contributions of all modes.

We start by writing up the Bell-states  encoded 
in the polarization degree of freedom of two photons with different momentum.
In terms of the creation operators of the input modes they take the form
\begin{eqnarray} 
\ket{\Psi^1}  & =  &\frac{1}{\sqrt{2}}\left( \ad_{1}  \ad_{3}   + 
\ad_{2}  \ad_{4}
\right) \ket{{\bf 0}} \label{eq:bella1} \\  
\ket{\Psi^2} &  = &\frac{1}{\sqrt{2}}\left( \ad_{1}  \ad_{3}   -
\ad_{2}  \ad_{4}
\right )
 \ket{{\bf 0}} \\  
\ket{\Psi^3} & =
&\frac{1}{\sqrt{2}}\left( \ad_{1} \ad_{4}  + \ad_{2}  \ad_{3} \right)  
\ket{{\bf 0}} \\ 
\ket{\Psi^4}   & = &\frac{1}{\sqrt{2}}\left( \ad_{1}  \ad_{4}   - \ad_{2}  \ad_{3}
\right) \ket{{\bf 0}}  
\label{eq:bella2}
\end{eqnarray} 
where $\ket{{\bf 0}}$ is the vacuum state, $\ad_{1}$ and $\ad_{2}$
 ($\ad_{3}$ and $\ad_{4}$)
correspond the two polarization modes of the first (second) photon.

Any two-photon state which enters the apparatus can be defined with 
 a bilinear form,
\begin{equation}
    \ket{\Psi}^{in} = \sum_{i,j=1}^n N_{i j}\ad_{i}\ad_{j} =
    {\bf a}\cdot{\bf N}\cdot{\bf a} \ket{{\bf 0}} 
    \label{eq:min}
\end{equation}
where ${\bf N}$ is a $ n \times n$ symmetric matrix and
${\bf a}=(a_{1}^\dagger,\ldots,a_{n}^\dagger)$  with 
the elements
$a_{i}^\dagger$ being the bosonic creation operators for the input modes.

Since the input and output modes are related through a linear 
transformation (Eq.~\ref{eq:linComb}) we can write a similar expression for 
the state in terms of the output  
modes $\{c_{i}^\dagger\}$,
\begin{equation}
     \ket{\Psi}^{in} =
    {\bf a}\cdot{\bf N}\cdot{\bf a} \ket{{\bf 0}}= {\bf c}\cdot U^T{\bf N} U \cdot {\bf c}={\bf 
    c}\cdot{\bf M}\cdot {\bf c}
\end{equation}
where ${\bf c}=(c_{1}^\dagger,\ldots,c_{n}^\dagger)$ and
\begin{equation}
{\bf M}=U^T {\bf N} U . 
    \label{eq:UMinU}
\end{equation}

Using the previous relation we can write the output bilinear form $M$
corresponding to each of the possible input Bell-States 
($\{\ket{\Psi^\mu}\}$) ,
\begin{equation}
    {\bf M}^\mu=U^T {\bf N}^\mu U  \;\; \mbox{where} \;\; 
    \mu=1,\ldots,4
    \label{eq:UMinBellU}
\end{equation}
and  from Eqs. (\ref{eq:bella1})-(\ref{eq:bella2}) the input bilinear forms are,
\begin{equation}
  {\bf N}^\mu=\frac{1}{2\sqrt{2}}\left( 
\begin{array}{c|ccc} 
{\bf W}^\mu & 0 &\cdots & 0  \\ 
\hline
0 &   &   &  \\
: &   & 0  & \\
0 &   &   & 
\end{array} 
\right). 
    \label{eq:N}
\end{equation}
where
\begin{equation}
  {\bf W}^\mu=\left( 
\begin{array}{cccc} 
0 & 0 & \delta_{\mu 1} + \delta_{\mu 2}& \delta_{\mu 3} +
\delta_{\mu 4}\\ 
0 & 0 & \delta_{\mu 3} - \delta_{\mu 4} & \delta_{\mu 1} - \delta_{\mu 2}\\ 
\delta_{\mu 1} + \delta_{\mu 2}&\delta_{\mu 3} - \delta_{\mu 4}& 0 & 0  \\ 
\delta_{\mu 3} + \delta_{\mu 4} & \delta_{\mu 1} - \delta_{\mu 2} &  
0 & 0 
\end{array} 
\right). \nonumber
    \label{eq:W}
\end{equation}

We can then rewrite Eq.~\eqref{eq:UMinBellU} as,
\begin{equation}
    {\bf M}^\mu=\frac{1}{2\sqrt{2}}U_{tr}^T {\bf W}^\mu U_{tr} 
    \label{eq:mout2}
\end{equation}
where $U_{tr}$ is the truncated version of $U$ obtained by taking only 
its first four rows (i.e. $\mbox{dim}(U_{tr})=4\times N$).
The matrices ${\bf W}^\mu$,
which are the first $4\times 4$ diagonal block of ${\bf N}^\mu$,
are unitary. This property is characteristic of all the maximally entangled states.

From the output bilinear forms ${\bf M}^\mu$ we can read out the contributions of 
the different input Bell states to particular detection events.

We start by studying the two-photon detection events.
The probability of having a two-photon detection at mode 
$c_{i}$  given the input state $\ket{\Psi}^{\mu}$ is,
\begin{equation}
      P_{i}^{\mu}[2]=\bra{0}c_{i}^2 M^{\mu^*}_{ii} M^\mu_{ii} 
       c_{i}^{\dagger^2}\ket{0}= 
      \frac{1}{4}|{\mathbf \alpha_{i}}\cdot W^\mu\cdot{\mathbf \alpha_{i}}|^2 \;\; 
      \label{eq:prb2}
\end{equation}
where ${\mathbf\alpha_{i}}$ is the $i^{th}$ column-vector of $U_{tr}$,
 i.e. ${\mathbf\alpha_{i}}=\{U_{1i},\ldots,U_{4i}\}$.
That is, the vector ${\mathbf \alpha_{i}}$ gives the linear relation between 
the output-mode $c_{i}^\dagger$ and the modes where we feed the input 
states ($a_{l}^\dagger,  l\leq 4$). 
Since the input modes are coupled to the auxiliary modes, the 
vectors $\mathbf\alpha_{i}$ are  in general not orthonormal. 

In order to identify unambiguously one Bell-State through a 
two-photon detection in mode $c_{i}$, the probability of this 
event  has to vanish for three of the 
Bell-States. Using Eq.~\eqref{eq:prb2} to impose this condition one can obtain the following solutions for ${\mathbf\alpha_{i}}$,
\begin{equation}
       {\mathbf\alpha_{i}}=(a,b,0,0) \;\; 
       \mbox{and}\;\;{\mathbf\alpha_{i}}=(0,0,a,b)
      \label{eq:alpha2ph}
\end{equation}
But it is easy to check that for both solutions the two-photon 
detection probability vanishes for all Bell-States, i.e 
$P[2]_{i}^{\mu}=0$ 
for $\mu=1,\ldots,4$.
So, a two-photon detection at 
given output mode $c_{i}$ cannot identify unambiguously 
one Bell state. After a two photon detection the remaining 
state is for all the Bell states the vacuum state, therefore an event of 
this type counts as a failure of the Bell-State analyzer.

Now we can concentrate on the single-photon detection event. 
From the result in the previous paragraph we
 know that these are the only events which can lead
 to a successful discrimination.
After a single photon detection in mode $c_{i}$ the remaining photon 
 is in the following conditional state,
\begin{equation}
    \ket{\Phi^{\mu}_{i}}=2 
    \sum_{j\neq i}^{N}M^{\mu}_{ij}c_{j}^{\dagger}\ket{{\bf 0}}= 
    2({\mathbf m_{i}^{\mu}}\cdot{\mathbf c}- M^{\mu}_{ii}c_{i}^{\dagger}) \ket{{\bf 0}}\; .
    \label{eq:psi}
\end{equation}
Again, the index $\mu$ stands for the different input Bell-States,
and ${\mathbf m_{i}^{\mu}}$ is the $i$th column vector of ${\mathbf M^{\mu}}$.
Using Eq.~\eqref{eq:mout2} we can write,
\begin{equation}
    {\mathbf m_{i}^{\mu}}=
    \frac{1}{2\sqrt{2}}U_{tr}^T {\bf W}^\mu {\mathbf\alpha_{i}} =
    \frac{1}{2\sqrt{2}}U_{tr}^T {\mathbf s_{i}}^\mu
    \label{eq:M1}
\end{equation}
where ${\bf s_{i}^\mu}={\bf W}^\mu {\bf \alpha_{i}}$.

It is straightforward to check that the vectors
$\{ {\mathbf s_{i}^1},\ldots,{\mathbf s_{i}^4}\}$,
 corresponding to the four input Bell states, are linearly dependent. 
This is formally expressed as
\begin{eqnarray}
   \mbox{det}({\mathbf s_{i}^1},\ldots,{\mathbf s_{i}^4})=0 &\Longrightarrow & \nonumber\\
  \sum_{\mu=1}^{4} b_{\mu} {\mathbf s_{i}^\mu}=0& &
 \mbox{ with at least one } 
    b_{\mu}\neq 0.
    \label{eq:lindeps}
\end{eqnarray}
Since they also have the same norm,
\begin{equation}
   |{\mathbf s_{i}^\mu}|^2 =  
   {\mathbf\alpha_{i}}^*{{\bf W}^\mu}^\dagger{\bf W}^\mu{\mathbf\alpha_{i}}=
  |{\mathbf \alpha_{i}}|^2  \mbox{ for } \mu=1,\ldots,4
    \label{eq:samenorm}
\end{equation}
at least two coefficients $b_{\mu}$ in 
Eq.~\eqref{eq:lindeps} must be non-zero.

From Eq.~\eqref{eq:lindeps} and  the linearity of
 Eq.~\eqref{eq:psi} and \eqref{eq:M1}
it follows that the conditional states after a one-photon detection 
are linearly dependent as well, 
\begin{equation}
   \sum_{\mu=1}^{4} b_{\mu} \ket{\Phi^{\mu}_{i}}=0 \;\;\mbox{ with at least two } 
    b_{\mu}\neq 0.
    \label{eq:lind}
\end{equation}
with the same coefficients
as the ${\mathbf s_{i}^\mu}$ dependence.
The overlaps of the conditional states are
\begin{equation}
    \braket{\Phi^{\eta}_{i}}{\Phi^{\mu}_{i}}=4
    ({\mathbf m_{i}}^{\eta^*}\cdot{\mathbf m_{i}}^{\mu}
    - M_{ii}^{\eta^*} M_{ii}^{\mu})
    \label{eq:psioverlap}
\end{equation}
or, by making use of Eq.~\eqref{eq:M1},
\begin{equation}
    \braket{\Phi^{\eta}_{i}}{\Phi^{\mu}_{i}}=\frac{1}{2}
    ({\mathbf s_{i}^{\eta}}^*\cdot{\mathbf 
    s_{i}^{\mu}}-
    ({\mathbf\alpha_{i}}.{\mathbf 
    s_{i}^{\eta}})^*({\mathbf\alpha_{i}}.{\mathbf s_{i}^{\mu}}))\; .
    \label{eq:psioverS}
\end{equation}

From the previous equation we can calculate the norm of the 
conditional states $\{\ket{\Phi^{\mu}_{i}}\}$ , i.e.
the probability of a one-photon detection at  mode $c_{i}$ for each input Bell-state,
\begin{eqnarray}
    P[1]_{i}^{\mu}&=& \braket{\Phi^{\mu}_{i}}{\Phi^{\mu}_{i}}=\frac{1}{2}
    (|{\mathbf s_{i}^{\mu}}|^2-
    |{\mathbf\alpha_{i}}.{\mathbf s_{i}^{\mu}}|^2) =
    \nonumber \\
    &=& \frac{1}{2}
     (|{\mathbf \alpha_{i}}|^2-
    |{\mathbf\alpha_{i}}.{\mathbf s_{i}^{\mu}}|^2)
    \label{eq:P1}
\end{eqnarray}

It is a well known fact \cite{chefles}
 that one cannot discriminate (not even with 
a small probability) states from a set of linearly dependent vectors.
But one should be careful when interpreting this statement.
What is true is that a state can  be discriminated from a set of vectors
 if and only if it is linearly independent of this set of vectors, i.e. 
it can not be written as a linear combination of the vectors in this set.
But, according to the definition in Eq.\eqref{eq:lind} it
might well be the case that a vector  $\ket{\Phi^{\nu}_{i}}$
from a linear dependent set is linearly independent from
the other vectors of the set. 
This would mean that the corresponding factor $b^\nu$ is zero.
A state with this property can then be discriminated from the 
rest with a non-zero probability. 
The fact that the four conditional states in Eq.\eqref{eq:lind} are 
linearly dependent with  at least  two non-zero
coefficients $b^\mu$ means that the minimum set of linear dependent states
contains two states. Consequently the 
maximum number of states which may be unambiguously discriminated,
 i.e. which are linearly independent of the others, is two.

Let us call $a$ and $b$  the  two different values of the index $\mu$ 
for which the conditional states are linearly independent. These correspond 
to the two Bell-states that with some probability
 may be unambiguously discriminated after a 
one photon detection in mode $c_{i}$. With this result and
 by assuming that the probability of unambiguous discrimination is one,
 we can put an upper bound to the \emph{success probability} $ p^s_{i}$
that the detector in mode $c_{i}$ is involved in the unambiguous 
discrimination of a Bell-State,
\begin{equation}
    p^s_{i}\leq \frac{1}{4}(P[1]_{i}^{a}+P[1]_{i}^{b})
    \leq\frac{1}{4}|{\mathbf \alpha_{i}}|^2
    \label{eq:Psuc}
\end{equation}
where the factor $\frac{1}{4}$ is the a priori probability 
of the corresponding initial Bell-states.

Each successful detection event involves single-photon detections in two 
different modes, say $c_i$ and $c_j$. This means that such an event contributes
to the corresponding probabilities of success, $p^s_{i}$ and $p^s_{i}$
  of both modes. This double counting is compensated by a factor $\frac{1}{2}$
 when collecting the contributions from all the modes to the total probability of success,
\begin{equation}
    P^s\leq\frac{1}{2}\sum_{i=1}^{n}p^s_{i}=
    \frac{1}{8}\sum_{i=1}^{n}|{\mathbf \alpha_{i}}|^2=\frac{1}{2}.
    \label{eq:PsucT}
\end{equation}
 Here we have made use of,
\begin{equation}
\sum_{i=1}^{n}|{\mathbf \alpha_{i}}|^2=\sum_{i=1}^{N}\sum_{j=1}^{4}
|{\mathbf U}_{ji}|^2=\sum_{j=1}^{4} 1 = 4 \; \mbox{.}
    \label{eq:sum}
\end{equation}

This upper bound can be achieved following the scheme proposed
 in \cite{weinfurter94}, and has been realized 
experimentally by Mattle \emph{et al.}\cite{mattle96} in  the implementation 
of the quantum dense coding scheme.
Using a 50/50 beam-splitter and two 
polarizing beam-splitters it is possible to unambiguously discriminate  
the states $\Psi^1$ and $\Psi^2$ successfully but $\Psi^3$ and $\Psi^4$ 
 give the same measurement outcomes.
One therefore obtains  a total probability of success of 
$P^s=\frac{1}{2}$. Since one can transform a Bell-state into any 
other Bell-state by local unitary transformations it is clear that 
one can unambiguously discriminate any other two Bell-states.
For example, by inserting a $\frac{\lambda}{2}$ plate before one of the 
input modes of the previous apparatus one would unambiguously 
discriminate the states $\Psi^3$ and $\Psi^4$. 
Not all linear-optical Bell-analyzers are restricted to detect two types of Bell states. For example, there exisist a fix array of beam splitters
 and phasers which unambiguously discriminates three Bell-states, but each 
of them with probability smaller than one so that $P^s\leq\frac{1}{2}$.

\section{Conclusions}

In this paper we have proven that the maximum probability of success
of a Bell-state analyzer build with a fix array of linear optical 
elements is $P^s=\frac{1}{2}$ for equiprobable input Bell-states.
Quantitative results for the  case
where conditional measurements and auxiliary photons can be used have 
not been given here, but a recent work by Knill, Laflamme and Milburn
\cite{knill}
shows that with these tools it is in principle possible to perform 
Bell measurements, among other crucial quantum measurements,
with a probability of failure arbitrarily small. The probability of failure
of their schemes decreases as $1/(n+1)$ with the number $n$ 
of highly entangled photons used in the auxiliary modes.
This makes their scheme unpracticable with the current technology.
 
Recently there have been some 
proposals to realize complete Bell measurements but all of them 
require non-linear optical elements \cite{paris,tomita} and
 are still far from being implementable with the current technology. 

With this paper we would like to stimulate the study of
 the capabilities and limitations of quantum information
 processing with linear elements 
when qubits are encoded in indistinguishable particles.
Only very recently some work has been done in this direction.
Bouwmeester \cite{bouwm00} has shown a way to implement a scheme which rejects
single bit-flip errors using linear-elements and 
an initial auxiliary GHZ state.
 Carollo \emph{et al.} \cite{carollo} have proved
the impossibility to realize a complete measurement in
 the ``non-local without entanglement'' basis \cite{bennett99}.

We acknowledge the Academy of Finland for financial support (project 43336).


\end{document}